\documentclass[aps,prb,twocolumn,superscriptaddress]{revtex4-1}

\usepackage[dvips]{graphicx}
\usepackage{color}
\usepackage[version=4]{mhchem}

\begin{document}

\title{Nanocriticality in the magnetic phase transition of CoO nanoparticles}

\author{Machteld E. Kamminga}
\affiliation{Nanoscience Center, Niels Bohr Institute, University of Copenhagen, 2100 Copenhagen {\O}, Denmark}
\author{Jonas Okkels Birk}
\affiliation{Nanoscience Center, Niels Bohr Institute, University of Copenhagen, 2100 Copenhagen {\O}, Denmark}
\affiliation{Currently at: Danish Technological Institute, Gregersensvej, 2630 Taastrup, Denmark}
\author{Jari {\'i} Hj\o llum}
\affiliation{Nanoscience Center, Niels Bohr Institute, University of Copenhagen, 2100 Copenhagen {\O}, Denmark}
\affiliation{Laboratory for Neutron Scattering and Imaging, Paul Scherrer Institut, 5232 Villigen, Switzerland}
\affiliation{Materials Research Department, National Laboratory for Sustainable Energy, Technical University of Denmark, 4000 Roskilde, Denmark}
\affiliation{Currently at: Faculty of Science and Technology, University of the Faroe Islands, 100 T{\'o}rshavn, Faroe Islands}
\author{Henrik Jacobsen}
\affiliation{Nanoscience Center, Niels Bohr Institute, University of Copenhagen, 2100 Copenhagen {\O}, Denmark}
\author{Jakob Lass}
\affiliation{Nanoscience Center, Niels Bohr Institute, University of Copenhagen, 2100 Copenhagen {\O}, Denmark}
\affiliation{Laboratory for Neutron Scattering and Imaging, Paul Scherrer Institut, 5232 Villigen, Switzerland}
\author{Thorbj\o rn L. Koch}
\affiliation{Nanoscience Center, Niels Bohr Institute, University of Copenhagen, 2100 Copenhagen {\O}, Denmark}
\affiliation{Currently at: BK Medical, Mileparken 34, 2730 Herlev, Denmark}
\author{Niels B. Christensen}
\affiliation{Department of Physics, Technical University of Denmark, 2800 Kgs.~Lyngby, Denmark}
\author{Christof Niedermayer}
\author{Lukas Keller}
\affiliation{Laboratory for Neutron Scattering and Imaging, Paul Scherrer Institut, 5232 Villigen, Switzerland}
\author{Luise Theil Kuhn}
\affiliation{Department of Energy Conversion and Storage, Technical University of Denmark, 2800 Kgs. Lyngby, Denmark}
\author{Elisabeth T. Ulrikkeholm}
\affiliation{Nanoscience Center, Niels Bohr Institute, University of Copenhagen, 2100 Copenhagen {\O}, Denmark}
\affiliation{Currently at: FOSS Analytical, Nils Foss All{\'e} 1, 3400 Hiller{\o}d, Denmark}
\author{Erik Brok}
\affiliation{Department of Physics, Technical University of Denmark, 2800 Kgs.~Lyngby, Denmark}
\affiliation{Currently at: FORCE Technology, Park All{\'e} 345, 2605 Br{\o}ndby, Denmark}
\author{Cathrine Frandsen}
\affiliation{Department of Physics, Technical University of Denmark, 2800 Kgs.~Lyngby, Denmark}
\author{Kim Lefmann}
\affiliation{Nanoscience Center, Niels Bohr Institute, University of Copenhagen, 2100 Copenhagen {\O}, Denmark}

\begin{abstract}
The universal theory of critical phase transitions describes the critical behavior at second-order phase transitions in infinitely large systems. With the increased contemporary interest in nanoscale materials, we investigated CoO nanoparticles by means of neutron scattering and found how the theory of critical phenomena breaks down in the nanoscale regime. Using CoO as a model system, we have identified a size-dependent nanocritical temperature region close to the antiferromagnetic phase transition where the magnetic correlation length of the nanoparticles converges to a constant value, which is significantly smaller than that of the saturated state found at low temperatures. This is in clear contrast to the divergence around $T_{\rm N}$ observed for bulk systems. Our findings of nanocriticality in the magnetic phase transition is of great importance for the understanding of phase transitions at the nanoscale.
\end{abstract}

\maketitle
%------------------------------------

\section{Introduction}
A key achievement of statistical mechanics in the last half of the 20th century is the description of phase transitions and critical phenomena, a universal behavior associated with second-order phase transitions which led to the development of a universal theory of critical phase transitions.\cite{kadanoff1967,griffiths1970} When a system is brought to a critical phase transition, many of its properties exhibit singular behavior.\cite{nishimori2011} In materials with magnetic phase transitions, the order parameter is the (sublattice) magnetization and for antiferromagnets, this spontaneous magnetization disappears at the N{\'e}el temperature, $T_{\rm N}$. While spins do not spontaneously (anti)align on the macroscopic scale at temperatures well above $T_{\rm N}$, the fluctuating spins remain correlated over a length scale, $\xi$ (the correlation length) which grows as $T_{\rm N}$ is approached.

The degree of singularity or divergence of physical quantities near the critical point is described by critical exponents.\cite{wilson1983} %For magnetic materials,
The system can be described by the {\em reduced temperature}, $t = (T - T_{\rm c})/T_{\rm c}$, with $T_{\rm c}$ the critical temperature, \textit{i.e.} $T_{\rm N}$ for antiferromagnets, and the order parameter follows power laws of $t$. For example, the antiferromagnetic order parameter, $M_{\rm AF}$, obeys $M_{\rm AF} \propto (-t)^\beta$ in close vicinity of $T_{\rm N}$, where $\beta$ is the {\em critical exponent} related to the magnetization. Above the critical temperature, the correlation length of the magnetic order follows a similar behavior, $\xi \propto t^{-\nu}$, where $\nu$ is the critical exponent of the correlation length. Similarly, the correlation length of the disordered domains in the ordered phase also follows a power law. Both correlation lengths diverge at the phase transition, indicating that both ordered and disordered phases percolate at $T_{\rm N}$, which is a key attribute of critical behavior.

Neutron scattering is well suited to study critical magnetic behavior, because it provides direct access to the values of the order parameter and correlation length. For example, antiferromagnetic order will lead to additional Bragg reflections in a neutron diffractogram with intensity $I\propto M_{\rm{AF}}^2=(-t)^{2\beta}$,
and the magnetic correlation length can be deduced from the width of the reflection.
Recent examples of critical magnetic scattering experiments include studies of the magnetic structure of \ce{MnBi2Te4}\cite{yan2019} and the magnetic phase transition of an artificial square ice system.\cite{sendetskyi2019} 

While the theory of critical phenomena is, strictly speaking, only valid for infinite systems, micrometer sized systems are sufficiently large to behave like an infinite system for all practical purposes. However, it is unclear to what degree nanoscale systems can be described by this theory, as the finite size of nanoparticles naturally prevents the correlation length from diverging. Nanoparticles have obtained increased attention in the last decade for a large variety of both medical and industrial applications.\cite{brigger2012,stark2015} These applications include, but are not limited to, batteries,\cite{sun2012,sun2012b,zhang2014,chen2019,wang2019} capacitors,\cite{zheng2014,bhattacharya2018} catalysis\cite{liao2014,zhao2015,jiang2016,kim2017,velegraki2018,park2018,zheng2018,park2019} and gas sensing.\cite{wang2018}  

In this work, we investigate the critical behavior in nanoparticles and study how the description of critical phenomena must be adjusted for phase transitions in the nanoscale regime. We use CoO as our model system, as CoO is a structurally simple Ising system and a relatively well-studied material in nanoparticle form.\cite{flipse1999,zhang2002,ghosh2005,ye2006,he2015,sun2012,sun2012b,zhang2014,chen2019,wang2019,zheng2014,bhattacharya2018,liao2014,zhao2015,jiang2016,kim2017,velegraki2018,park2018,zheng2018,park2019,wang2018} Moreover, despite the frustration inherent to its face-centered cubic structure, bulk CoO has a second-order antiferromagnetic phase transition near room temperature (with a critical temperature, $T_{\rm N} \approx 289$~K).\cite{rechtin1971b} As $T_{\rm N}$ is close to room temperature, nanoparticles of CoO can be studied in the temperature region near $T_{\rm N}$ without the samples being destroyed by heating. In the antiferromagnetic phase, the magnetic structure is given by alternating planes of ferromagnetically aligned spins that align antiferromagnetically along the (1 1 1) direction, similar to MnO, FeO and NiO.\cite{roth1958} Moreover, critical magnetic neutron scattering experiments on bulk single crystals of CoO have been performed about half a century ago.\cite{mcreynolds1959,rechtin1971}

By means of neutron scattering, we show that in contrast to micrometersized CoO, the theory of critical phenomena breaks down for CoO nanoparticles.  Furthermore, we qualitatively support our experimental observation of this nanocritical behavior by Monte Carlo simulations. Our findings provide an additional branch to the theory of critical phenomena, that is important to the understanding of magnetic phase transitions in nanosized or confined systems.

%------------------------------------

\section{Experimental}
The CoO nanoparticles were prepared by a method similar to one previously reported.\cite{frandsen2004} Here, \ce{(CH3COO)2Co} $\cdot$ \ce{4H2O} was suspended in ethanol, baked at 100 $^{\circ}$C and consecutively annealed at a temperature between 325 $^{\circ}$C and 425 $^{\circ}$C under constant argon flow to remove acetic acid and water. By increasing the temperature and annealing time, the particles were allowed to grow together, resulting in larger-sized particles. Three CoO nanoparticle samples were obtained with nominal diameters of 20 nm, 30 nm, and 40 nm.  The samples were characterized by X-ray diffraction on a Rikagu rotating anode using Cu-K$\alpha$ radiation with a wavelength $\lambda = 1.54$~\AA. As shown in the Supplemental Material (SM),\cite{supplementary} the crystalline size of nanoparticles were determined to be 21.3(8) nm, 29.3(4) nm and 41.7(5) nm, respectively. For comparison, a sample of micrometer-sized CoO particles was commercially purchased.

Neutron diffraction was carried out at the Paul Scherrer Institute (CH),\cite{PSI} using the \mbox{RITA-2} cold-neutron spectrometer \cite{bahl2004} in two-axis mode with a wavelength of 4.7~\AA, taking advantage of the large area of the position-sensitive detector. Additional diffraction data were taken at the cold-neutron powder diffractometer DMC at the Paul Scherrer Institute, using a wavelength of 4.2~\AA~{} and the full detector bank covering $80^\circ$ of scattering angles.
We fitted the magnetic peaks using a Voigt function, where the Lorentzian half width at half maximum ($\Gamma$) is the broadening caused by the finite size of correlated domains, and the Gaussian width is the resolution of the instrument, determined by fitting the data obtained on the micrometer-sized particles at 10 K. We also allowed for a sloping background in the fitting.

Unfortunately, the 20 nm data appeared to be of insufficient quality to perform proper data fitting and are therefore excluded from further analysis in this work (see Fig. S2 in the SM for an example of insufficient data quality \cite{supplementary}). We also note that the work of Ghosh \textit{et al.} demonstrated that very small CoO nanoparticles ($<$ 16 nm) contain different magnetic behavior (ferromagnetic interactions) than their larger counterparts, putting a lower limit to the size of antiferromagnetic CoO nanoparticles.\cite{ghosh2005}

%------------------------------------

\section{Results and discussion}
\begin{figure}[b]
\includegraphics[width=0.45\textwidth]{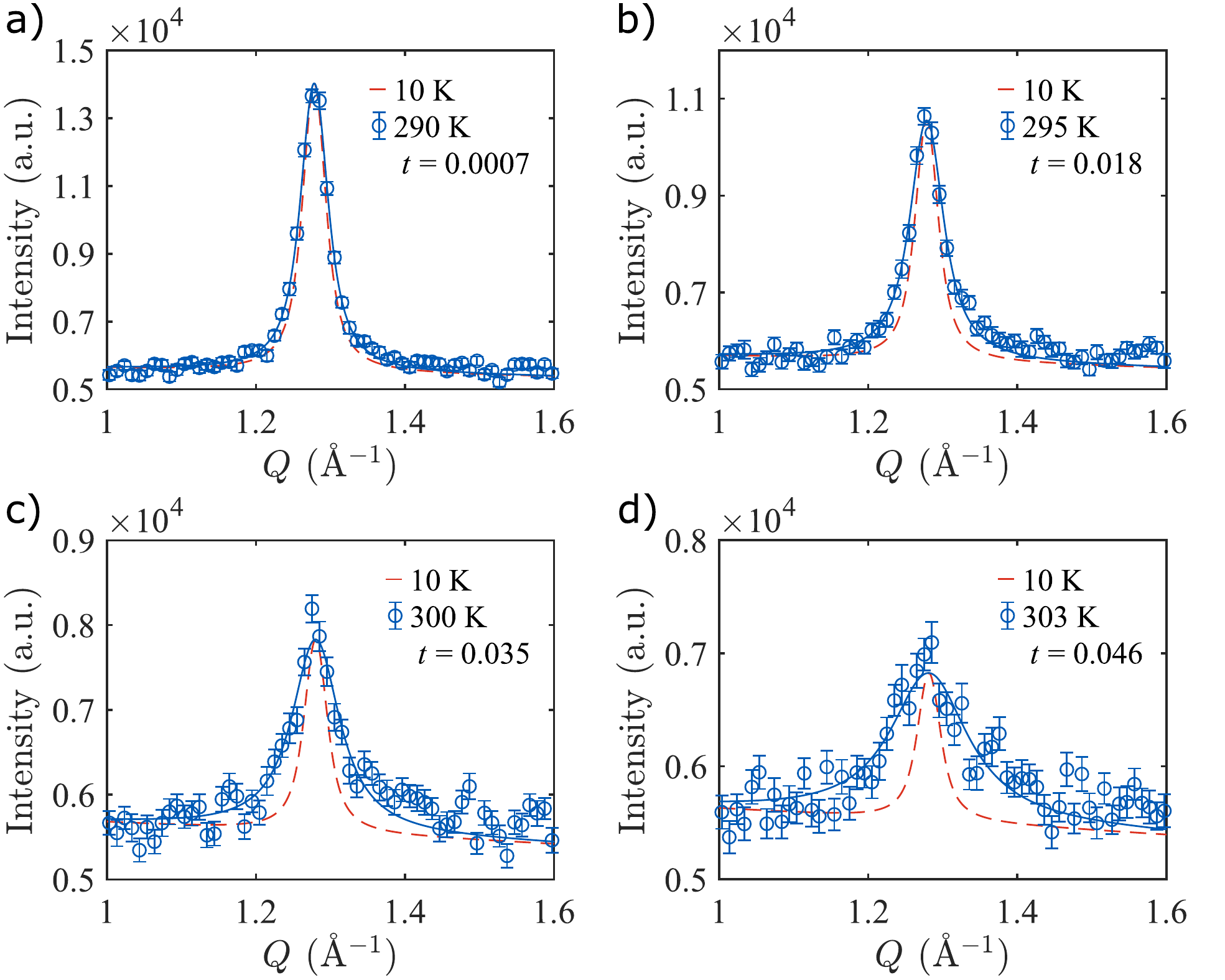}
\caption{Development of critical scattering near the antiferromagnetic phase transition in 30 nm CoO particles, at \textbf{a)} 290~K, \textbf{b)} 295~K, \textbf{c)} 300~K and \textbf{d)} 303~K. The measured neutron diffraction intensities as a function of momentum transfer are depicted in blue (the error bars represent one standard deviation), the fit to the data is given by the continuous blue line (Voigt) and the red dashed line represents the fit to the long-range ordered signal measured at 10 K, as given in Fig. S3 in the SM,\cite{supplementary} but re-scaled to match the intensity of each peak.} 
\label{Fig:criticalmethod}
\end{figure}

In our neutron experiment, we observed the critical magnetic scattering around the (1/2 1/2 1/2) peak. Typical results of the development of this critical scattering in the 30 nm sample for four temperatures above $T_{\rm N}$ is shown in Fig.~\ref{Fig:criticalmethod}. The 40 nm sample shows very similar behavior, as will become apparent below. It was not possible to measure the critical scattering at temperatures below $T_{\rm N}$, as the very strong signal from the magnetic Bragg peak overshadows the signal of the critical scattering at lower temperatures. This is in contrast to the work by Rechtin and Averbach, who could easily separate the magnetic Bragg-scattering component in their study on a single crystal with sharp collimation.\cite{rechtin1971} The fact that we use nanoparticle samples in our study significantly enhances the complexity of our data analysis. However, we were able to investigate the temperature dependence of the magnetic correlation length, and its critical exponent $\nu$, at the phase transition as approached from the high-temperature side.

%=================== FIGURE 1 ==================================
%=================== FIGURE 1 ==================================

The magnetic correlation length, $\xi$, is inversely proportional to the width of the magnetic diffraction peak: $\xi = 1 / \Gamma$, where $\Gamma$ is the half width at half maximum (HWHM). 
As shown in Fig.~\ref{Fig:criticalmethod}, the width of the critical scattering increases with temperature. This means that the size of magnetic domains, measured as magnetic correlation length, is largest at $T_{\rm N}$ and decreases for increasing temperature, in agreement with the theory of critical phenomena as established for infinitely large systems. However, we observe that the critical scattering as measured very close to $T_{\rm N}$ (see Fig.~\ref{Fig:criticalmethod}\textbf{a}) is slightly broader than the Bragg scattering measured at 10 K. This low-temperature data corresponds to the completely ordered magnetic structure and the width of the peak is given by the resolution of the instrument and the finite size of the magnetic structure of the nanoparticles. Thus, as expected, the magnetic correlation length appears to not diverge at $T_{\rm N}$ for CoO nanoparticles, as it does for infinite systems. 

In order to analyze the critical scattering data as a function of temperature, it is crucial to precisely determine the critical temperature for each sample, as $T_{\rm N}$ might depend on the particle size. For single-crystal or micrometer-sized samples, $T_{\rm N}$ is usually determined by following the sharp magnetic Bragg peak below $T_{\rm N}$, fitting the peak intensity in the approximate range $0.1 < -t < 0.01$, with $t = (T - T_{\rm N})/T_{\rm N}$, where we expect the peak intensity to scale as $ \propto (-t)^{2\beta}$. However, for the nanoparticles this procedure proved unreliable due to the relatively higher background levels and finite-size broadening of the Bragg peak, giving unacceptable uncertainties of $T_{\rm N}$ of the order 3-4 degrees. Instead, we found that for nanoparticles, a much better method was to determine $T_{\rm N}$ from the high-temperature data. Here, we expect a power-law behaviour of the form $\xi \propto t^{-\nu}$, and $T_N$ was defined as the value where the high-temperature data best follows this equation, using a simple $\chi^2$ fit. 

\begin{figure}[t]
%\centering
\includegraphics[width=0.45\textwidth]{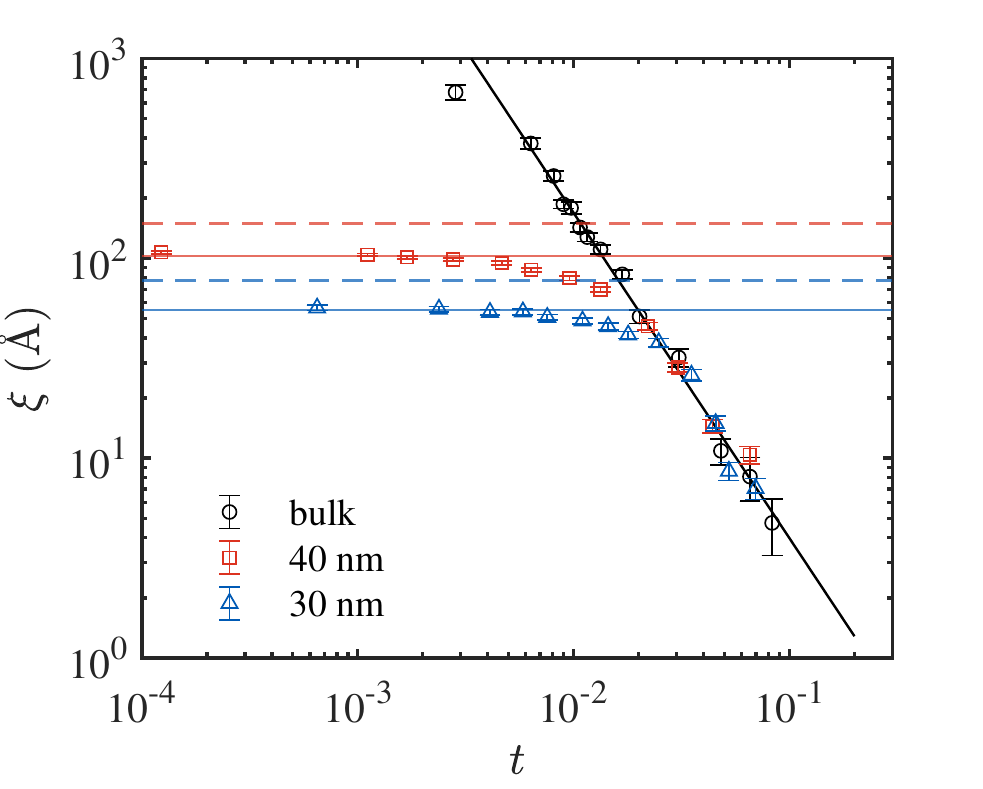}
\caption{Correlation length of short-range order for 30 and 40 nm CoO particles as a function of reduced temperature, $t = (T - T_{\rm N})/T_{\rm N}$. The black line corresponds to the power-law fit to the bulk CoO data. The solid blue and red lines correspond to the converged values of the correlation lengths close to $T_{\rm N}$ for the 30 and 40 nm particles, respectively. The dashed lines denote the correlation length of the corresponding long-range ordered state, measured at 10 K.}
\label{Fig:correlationreducedt}
\end{figure}

For the CoO bulk sample, \textit{i.e.} the commercially-bought micrometer-sized CoO, we performed this power-law analysis of the correlation length, resulting in a $T_{\rm N}$ of 286.2(4) K and a critical exponent of $\nu = 1.6(4)$. As our obtained value of $\nu$ is higher than 0.63, expected for a 3D Ising system,\cite{Guida1998} we argue that the CoO system appears to be more complex and cannot be described by such a simple model. In fact, previous work has shown that the critical exponent $\beta$ (see SM\cite{supplementary}) also does not follow the simple 3D Ising model as it is severely affected by a tetragonal lattice contraction below $T_{\rm N}$. \cite{Rechtin1970} We note, however, that to the best of our knowledge, we are the first to report experimental values of $\nu$ for CoO and further assessment of the origin of this variation in critical exponents is beyond the scope of this work. The value of $T_{\rm N}$ is in agreement with the value found by fitting the peak intensity as explained above. Note, however, that this deviates a bit from the expected value of $T_{\rm N} \approx 289$~K,\cite{rechtin1971b}. The reason for this deviation is that the cryostats are calibrated for low temperatures and not for high temperatures, leading to a small offset in the apparent temperature near room temperature. 
In addition, the samples were not all measured in the same cryostat, and so the offset might be different for different samples.
However, this would not affect any relative temperature differences between the measurements on the same sample. For the 30 nm and 40 nm samples, $T_{\rm N}$ was determined by fitting the highest temperature data to the same model as described for the bulk data, using the least-squares method. We argue that this method is accurate as in the limit of the smallest domain sizes, the size of the particle itself does not matter, and all data should therefore lie on the same line (\textit{i.e.}, the black line shown in  Fig.~\ref{Fig:correlationreducedt}). This method yielded a $T_{\rm N}$ of 289.8(5) K and 290.2(4) K for the 30 nm and 40 nm samples, respectively. Additionally, these N\'eel temperatures were used to determine the critical exponent, $\beta$, of the magnetic order parameter, as shown in Fig. S4 in the SM.\cite{supplementary}

Fig.~\ref{Fig:correlationreducedt} shows the measured correlation length as a function of reduced temperature for the CoO nanoparticles in comparison to the bulk data. The bulk data clearly shows a divergence of the correlation length near $T_{\rm N}$, as expected by the universal theory of critical phase transitions.\cite{kadanoff1967,griffiths1970} However, two different regions are apparent for the nanoparticle data: 1) the region above $t = 0.03$ where it follows the bulk data, and 2) a converging correlation length below reduced temperatures of $t = 0.01$, corresponding to temperatures between $T_{\rm N}$ and $\approx 292$~K. Thus, as expected, no divergence of the correlation length near $T_{\rm N}$ is observed for the nanoparticle samples. In fact, the value of the converged correlation length depends on the size of the nanoparticles: 55(2) \AA{} and 103(2) \AA{} for the 30 nm and 40 nm particles, respectively. These converged values correspond to roughly 2/3 of that of the long-range ordered state measured at 10
K (77(2) \AA{} and 149(2) \AA{}, for the 30 nm and 40 nm particles, respectively).  

%=================== FIGURE 2 ==================================
%=================== FIGURE 2 ==================================

Note that the correlation length of the magnetic domains determined as $\xi=1/\Gamma$ is not the same entity as the diameter ($D$) of the ordered magnetic regions in the nanoparticles at low temperatures. The latter can be calculated using the Scherrer equation, $D=\pi K/\Gamma$, where $K$ is a dimensionless shape factor. Using Scherrer's value of $K=0.94$,\cite{Scherrer1918} we find  $D=21.3(5)$ nm and $D=44.0(5)$ nm at 10 K for the 30 and 40 nm particles, respectively. These values indicate that there are magnetic dead layers on the surface of the 30 nm particles, as seen in other nanoparticles \cite{Curiale2009}, while the 40 nm particles are fully ordered. We now focus on the correlation length of the magnetic domains near $T_{\rm N}$.

It is logical that no infinitely large correlation lengths can be observed in finite systems; the size of the nanoparticles already provides an upper limit. However, for both sizes of nanoparticles, the converged value of the correlation length only correspond to about 2/3 of that of the long-range ordered state as measured at 10 K, which is the longest correlation length observed in the particles. This means that no true long-range order exist in the nanoparticles near the phase transition. We therefore conclude that magnetically ordered and disordered domains coexist over a region of a few degrees around $T_{\rm N}$, in what appears to be a semistable equilibrium.

%--------------------------- Simulations-------------------------------------

To investigate the behavior of magnetic nanoparticles near $T_{\rm N}$ in more detail, we carried out classical Monte Carlo simulations using a simple nearest neighbor Ising model on a cubic lattice with a lattice constant equal to the Co-Co distance in CoO, $a=4.2615/\sqrt(2)$ \AA{}. To capture the physics of spherical, monodisperse nanoparticles, we used open boundary conditions and included only spins within a sphere of diameter $D$. We used the Metropolis Hastings algorithm\cite{metropolis1953,hastings1970}, where spin flips that reduce the energy were always kept, and spin flips that increase the energy were kept with a probability of $\exp[{-\Delta E/T_s}]$, where $\Delta E$ is the change in energy and $T_s$ is the simulated temperature of the system. The phase transition was found at $T_s\approx 4.5$. The simulations were carried out on the Quantum Wolf Cluster at the Laboratory for Quantum Magnetism, EPFL, Switzerland. More details regarding the simulations are given in the SM.\cite{supplementary}

%=================== FIGURE 3 ==================================
\begin{figure}[t]
\includegraphics[width=0.23\textwidth]{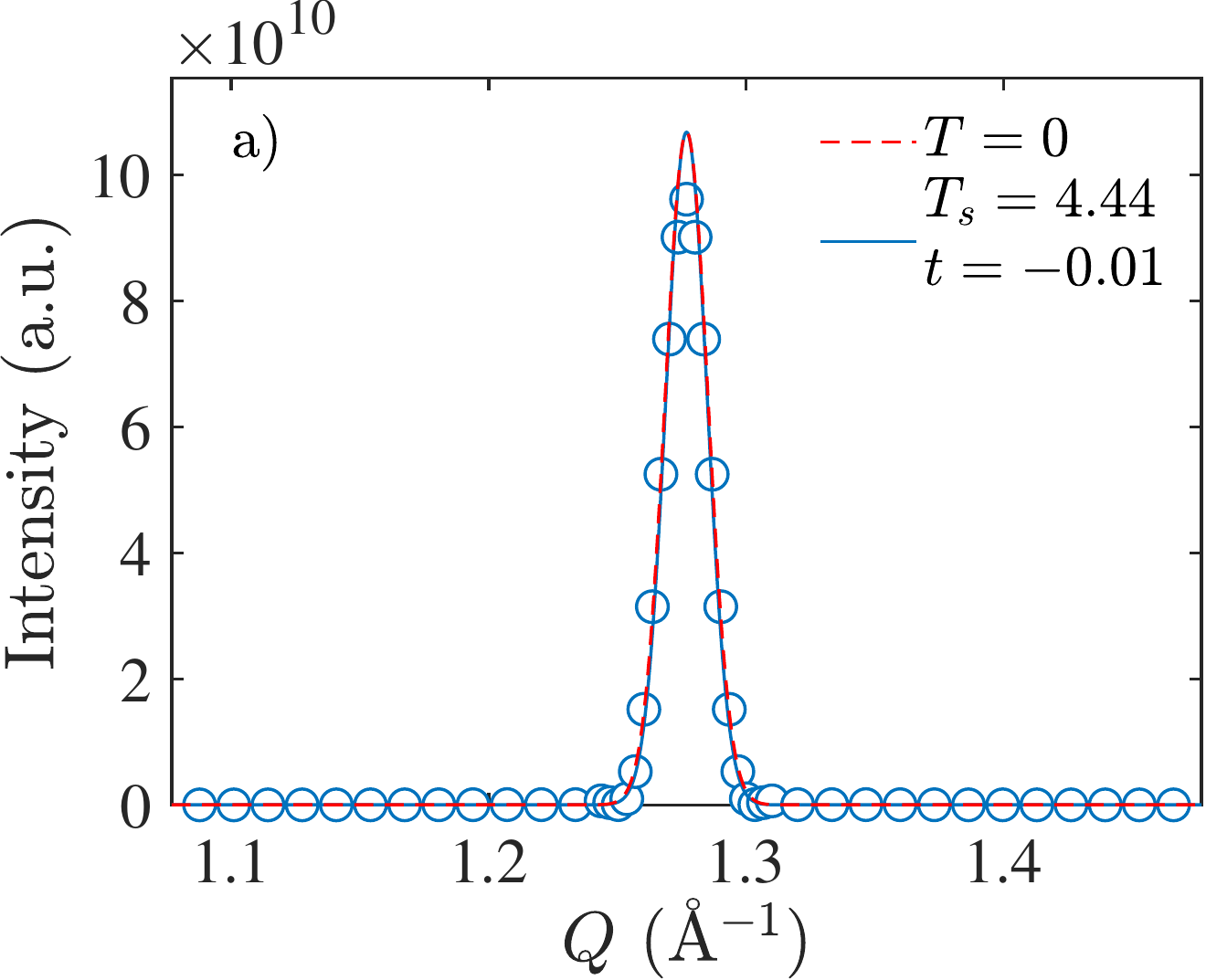}
\includegraphics[width=0.23\textwidth]{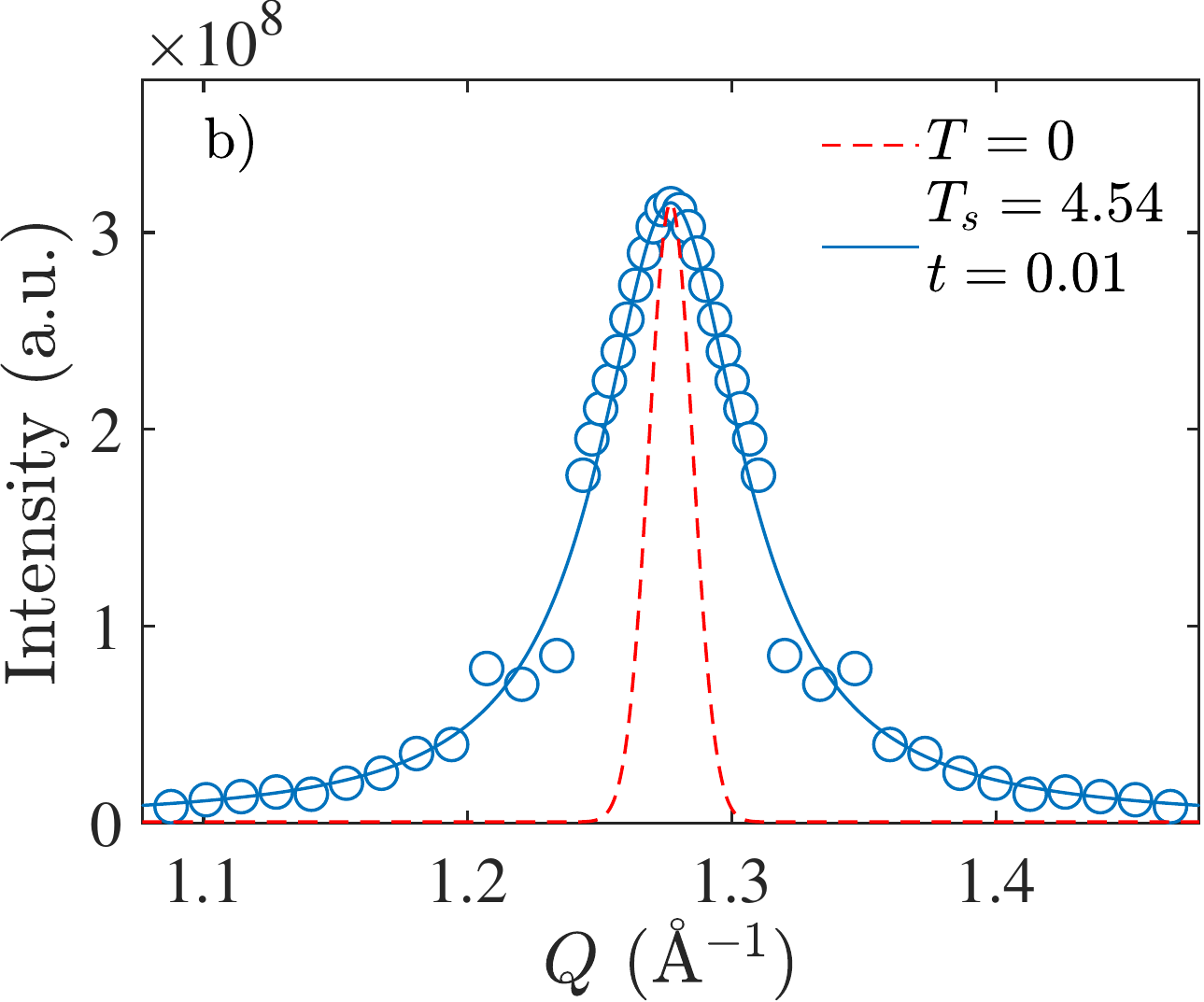}
\caption{Simulated scattering \textbf{a)} below ($t=-0.01$) and \textbf{b)} above ($t=0.01$) the magnetic phase transition in a magnetic nanoparticle with a diameter of 36 nm. The fit to the data is shown by the continuous blue line and the red dashed line represents the fit to the long-range ordered signal at base temperature, but re-scaled to match the intensity of each peak. }
\label{Fig:montecarlo_fits}
\end{figure}
%=================== FIGURE 3 ==================================

%=================== FIGURE 4 ==================================
\begin{figure}[b]
%\centering
\includegraphics[width=0.45\textwidth]{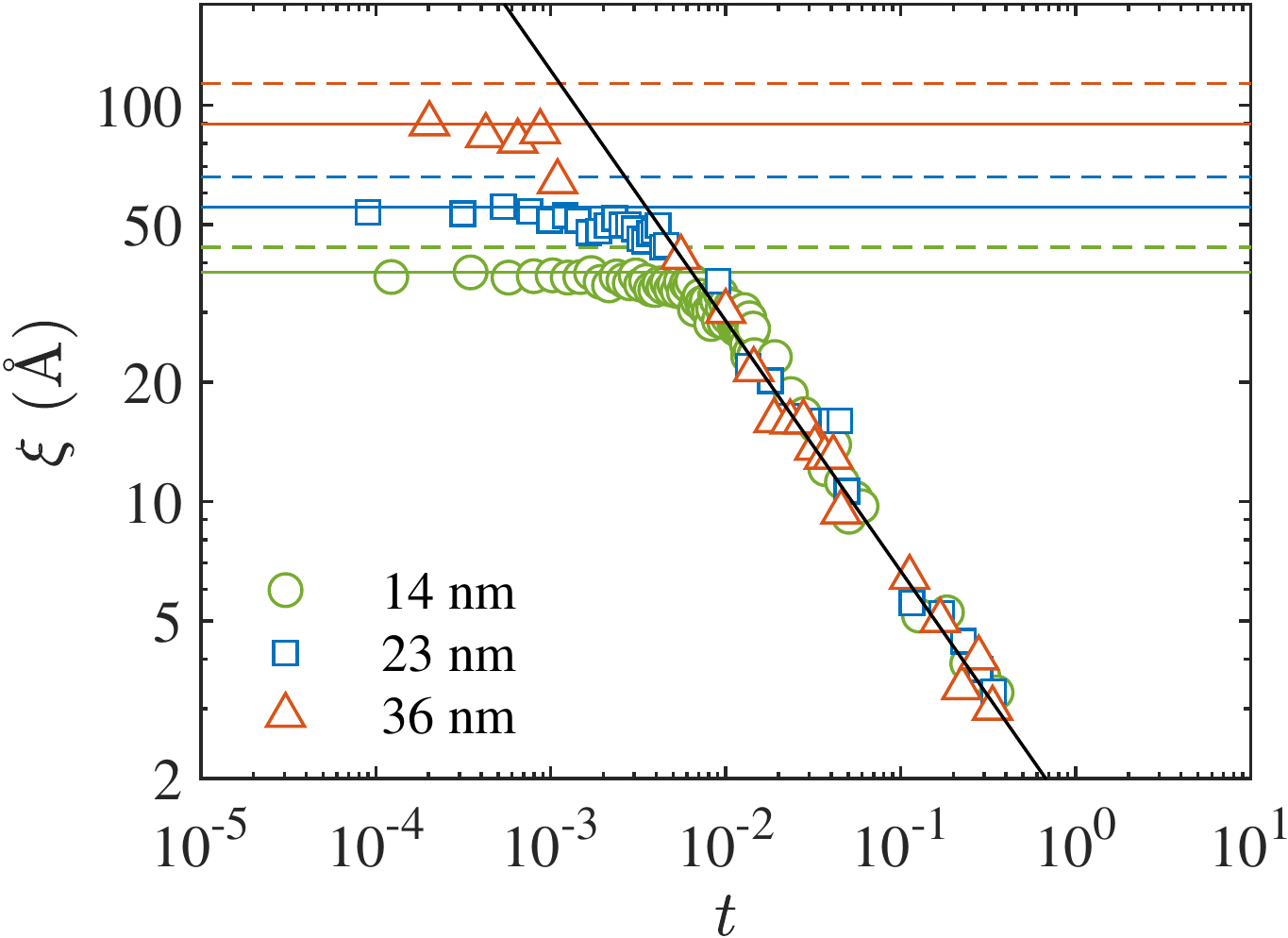}
\caption{Correlation length of short-range order for simulated magnetic nanoparticles of different sizes as a function of reduced temperature. The black line corresponds to the expected bulk behavior with $\nu=0.63$. The solid colored lines correspond to the converged values of the correlation lengths close to $T_{\rm N}$ for the nanoparticles of different sizes. The dashed lines denote the correlation length of the corresponding long-range ordered state, obtained at $T = 0$. 
}
\label{Fig:montecarlo_xi}
\end{figure}

%=================== FIGURE 4 ==================================

For consistency, we analyzed the simulated data using the same approach as for the experimental data. At the lowest temperatures, the intensity is well approximated by a Gaussian, which captures the finite size of the particles. At higher temperatures, the signal broadens and can be described by a Voigt function, in which the Lorentzian part accounts for the additional broadening. Examples of the simulated signal of a 36 nm particle below and above the phase transition are shown in Fig.~\ref{Fig:montecarlo_fits}. 

Fig.~\ref{Fig:montecarlo_xi} shows the correlation length for three nanoparticle diameters as function of reduced temperature. As for the experimental data, we calculated the correlation length using $\xi=1/\Gamma$, where $\Gamma$ is the HWHM of the signal. There is a clear qualitative agreement between our experimental data and simulations. The simulations show two different regions, one where the correlation length follows the expected bulk behavior and one where it converges to a constant value, indicating a lack of divergence of the correlation length near the critical temperature in the nanoscale regime. Moreover, this value of the converged correlation length depends on the size of the nanoparticles and only reaches a fraction of that of the long-range ordered state obtained at $T = 0$. This is in close agreement with the experimental data. Note that a quantitative comparison between the experimental and simulated data is compromised by the simplified simple cubic lattice used in the simulations, which does not include the magnetic frustration present in the CoO lattice. For example, the critical exponent of the simulated bulk behavior ($\nu = 0.63(3)$, corresponding to a 3D Ising system\cite{Guida1998}) deviates from that of the experimental data ($\nu = 1.6(4)$), as discussed above.

%------------------------------------

\section{Conclusions}
In conclusion, we used neutron scattering to measure the critical magnetic scattering near the antiferromagentic phase transition in CoO at the nanoscale. Our results show that nanoparticles of CoO exhibit a different critical scattering behavior at temperatures close to $T_{\rm N}$, as compared to their bulk counterpart. In contrast to the divergence in correlation length observed for larger systems, a converged value of the correlation length close to the phase transition is observed at the nanoscale. Notably, the converged value is significantly smaller than that of the saturated state observed at low temperatures. Moreover, the size of the maximum correlation length depends on the size of the nanoparticles. Our Monte Carlo simulations support our findings of such a converged correlation length near the phase transition. We hereby show that the theory of critical phenomena, developed for macroscopic systems displaying continuous phase transitions, requires modifications when applied to a nanoscale system in which geometrical constraints on the correlation need to be taken into account. We emphasize that while our study deals with magnetic nanoparticles, such modifications would be required for any nanoscale system.
%------------------------------------

\section*{Acknowledgments}
We thank N.H. Andersen, B. Lebech, P.-A. Lindg\aa rd, J. Juul, and H. Bruus for stimulating discussions. A large thank goes to  S.~M\o rup for participating in the initial phases of this project. We thank H.M. R\o nnow for providing access to the Quantum Wolf computer cluster at the Laboratory for Quantum Magnetism, EPFL, Lausanne. MEK was supported by MSCA-IF Horizon 2020, grant number 838926. HJ acknowledges support from the Carlsberg Foundation. This work was supported by the Danish Technical Research Council through the Nanomagnetism framework program, and the Danish Natural Science Research Council through DANSCATT. The work is based on experiments performed at SINQ, Paul Scherrer Institute, Switzerland.

MEK and JOB contributed equally to this work.

%------------------------------------

\end{document}

\title{Supplemental Material:\\ Nanocriticality in the magnetic phase transition of CoO nanoparticles}

\author{Machteld E. Kamminga}
\affiliation{Nanoscience Center, Niels Bohr Institute, University of Copenhagen, 2100 Copenhagen {\O}, Denmark}
\author{Jonas Okkels Birk}
\affiliation{Nanoscience Center, Niels Bohr Institute, University of Copenhagen, 2100 Copenhagen {\O}, Denmark}
\affiliation{Currently at: Danish Technological Institute, Gregersensvej, 2630 Taastrup, Denmark}
\author{Jari {\'i} Hj\o llum}
\affiliation{Nanoscience Center, Niels Bohr Institute, University of Copenhagen, 2100 Copenhagen {\O}, Denmark}
\affiliation{Laboratory for Neutron Scattering and Imaging, Paul Scherrer Institut, 5232 Villigen, Switzerland}
\affiliation{Materials Research Department, National Laboratory for Sustainable Energy, Technical University of Denmark, 4000 Roskilde, Denmark}
\affiliation{Currently at: Faculty of Science and Technology, University of the Faroe Islands, 100 T{\'o}rshavn, Faroe Islands}
\author{Henrik Jacobsen}
\affiliation{Nanoscience Center, Niels Bohr Institute, University of Copenhagen, 2100 Copenhagen {\O}, Denmark}
\author{Jakob Lass}
\affiliation{Nanoscience Center, Niels Bohr Institute, University of Copenhagen, 2100 Copenhagen {\O}, Denmark}
\affiliation{Laboratory for Neutron Scattering and Imaging, Paul Scherrer Institut, 5232 Villigen, Switzerland}
\author{Thorbj\o rn L. Koch}
\affiliation{Nanoscience Center, Niels Bohr Institute, University of Copenhagen, 2100 Copenhagen {\O}, Denmark}
\affiliation{Currently at: BK Medical, Mileparken 34, 2730 Herlev, Denmark}\author{Niels B. Christensen}
\affiliation{Department of Physics, Technical University of Denmark, 2800 Kgs.~Lyngby, Denmark}
\author{Christof Niedermayer}
\author{Lukas Keller}
\affiliation{Laboratory for Neutron Scattering and Imaging, Paul Scherrer Institut, 5232 Villigen, Switzerland}
\author{Luise Theil Kuhn}
\affiliation{Department of Energy Conversion and Storage, Technical University of Denmark, 2800 Kgs. Lyngby, Denmark}
\author{Elisabeth T. Ulrikkeholm}
\affiliation{Nanoscience Center, Niels Bohr Institute, University of Copenhagen, 2100 Copenhagen {\O}, Denmark}
\affiliation{Currently at: FOSS Analytical, Nils Foss All{\'e} 1, 3400 Hiller{\o}d, Denmark}
\author{Erik Brok}
\affiliation{Department of Physics, Technical University of Denmark, 2800 Kgs.~Lyngby, Denmark}
\affiliation{Currently at: FORCE Technology, Park All{\'e} 345, 2605 Br{\o}ndby, Denmark}
\author{Cathrine Frandsen}
\affiliation{Department of Physics, Technical University of Denmark, 2800 Kgs.~Lyngby, Denmark}
\author{Kim Lefmann}
\affiliation{Nanoscience Center, Niels Bohr Institute, University of Copenhagen, 2100 Copenhagen {\O}, Denmark}

\maketitle

%=================== FIGURE S1 ==================================
\begin{figure}
\centering
\includegraphics[width=0.28\textwidth]{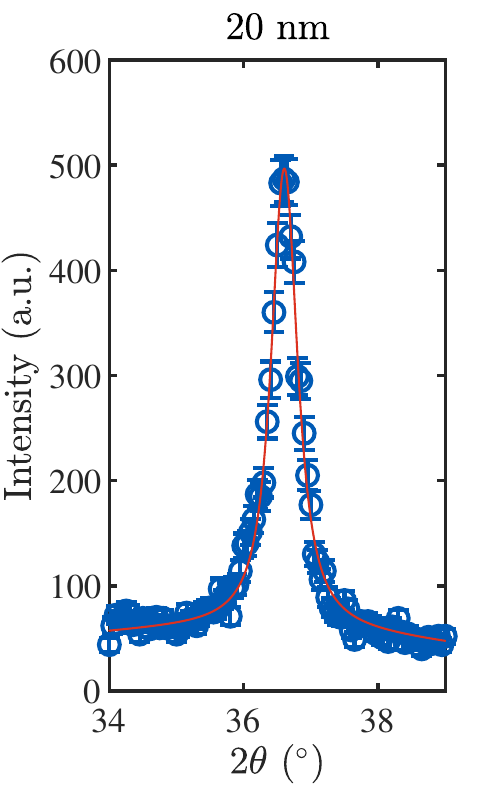}
\includegraphics[width=0.28\textwidth]{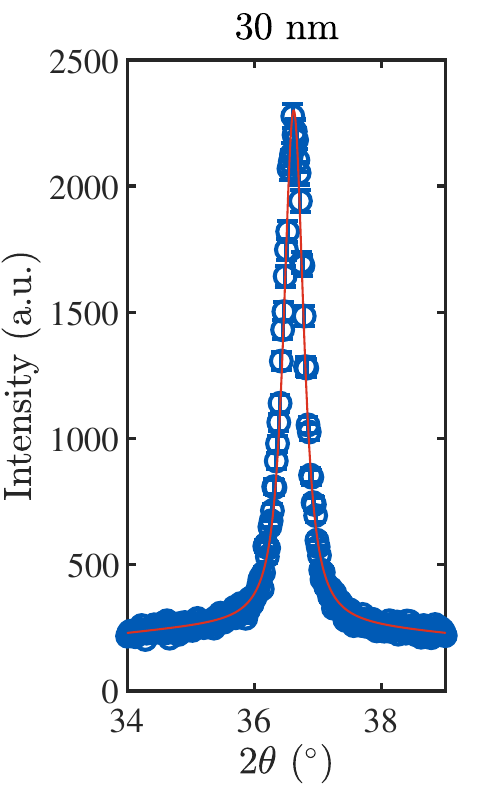}
\includegraphics[width=0.28\textwidth]{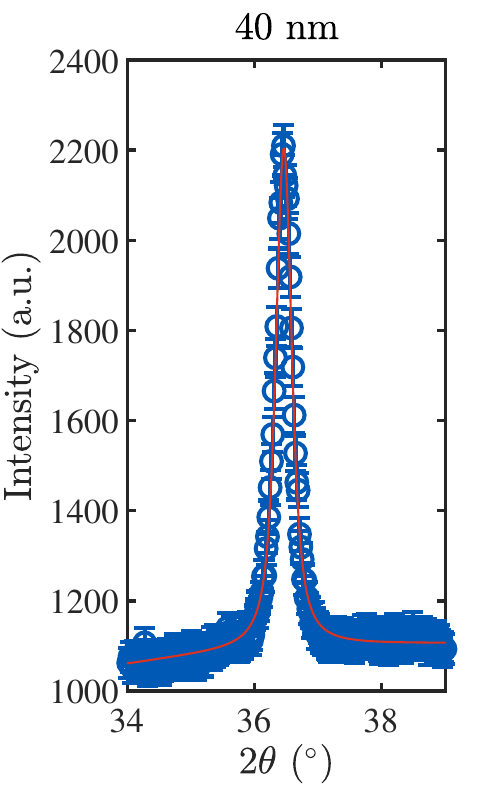}
\caption{X-ray powder diffraction measurements on the nano-sized CoO samples. The characteristic structural (1 1 1) reflection
at $Q = 2.56$~\AA$^{-1}$ is clearly visible for all samples. As expected, the peak widths are inversely proportional to the size of the nanoparticles. Using the Scherrer equation $\tau = K \lambda / \beta \cos\theta$, with $K=0.94$ the dimensionless shape factor,\cite{Scherrer1918} $\lambda$ the X-ray wavelength (1.5406 \AA), $\beta$ the line broadening (full width at half maximum) and $\theta$ the Bragg angle, the particle sizes ($\tau$) are determined to be 21.3(8) nm, 29.3(4) nm and 41.7(5) nm, respectively.}
\label{Fig:xRay}
\end{figure}
%=================== FIGURE S1 ==================================

%=================== FIGURE S2 ==================================
\begin{figure}
%\centering
\includegraphics[width=0.5\textwidth]{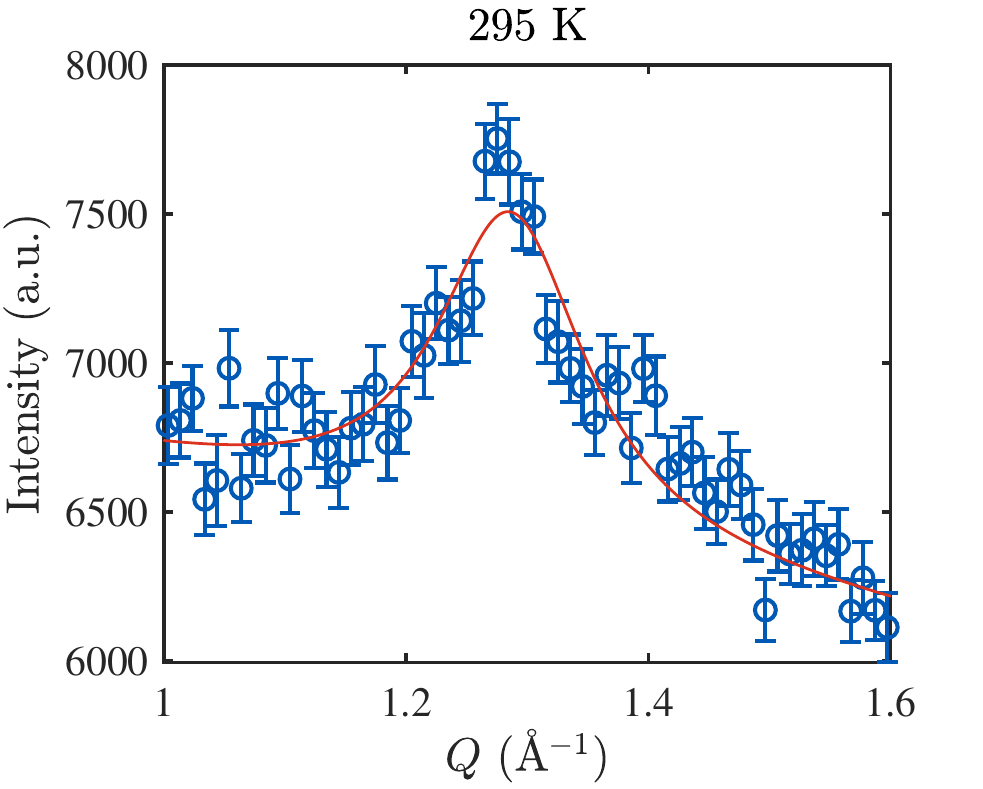}
\caption{Example of insufficient data quality for the 20 nm sample. No proper fit to this neutron scattering data could be obtained.}
\label{Fig:2Kscattering}
\end{figure}
%=================== FIGURE S2 ==================================

%=================== FIGURE S3 ==================================
\begin{figure}
%\centering
\includegraphics[width=0.5\textwidth]{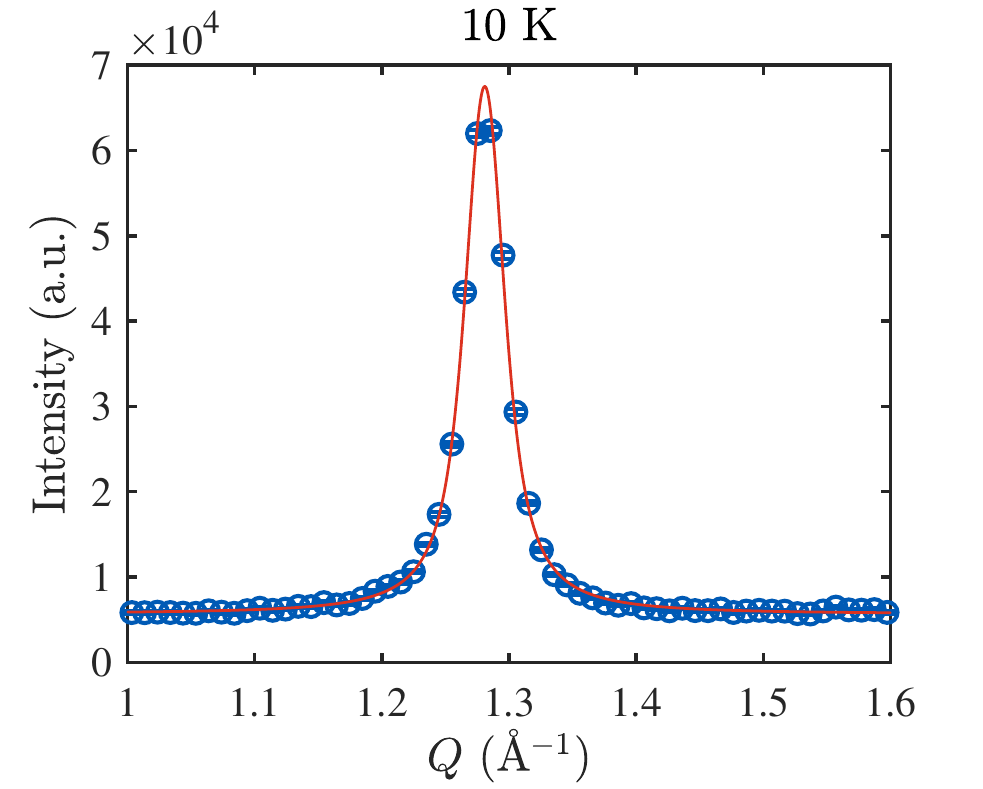}
\caption{Scattering from 30 nm CoO particles at 10 K (long-range ordered signal). The measured neutron diffraction intensity as a function of momentum transfer is depicted in blue, and a fit to the data is given by the continuous red line (Voigt).}
\label{Fig:10Kscattering}
\end{figure}
%=================== FIGURE S3 ==================================

%=================== FIGURE S4 ==================================
\begin{figure}
\centering
\includegraphics[width=0.48\textwidth]{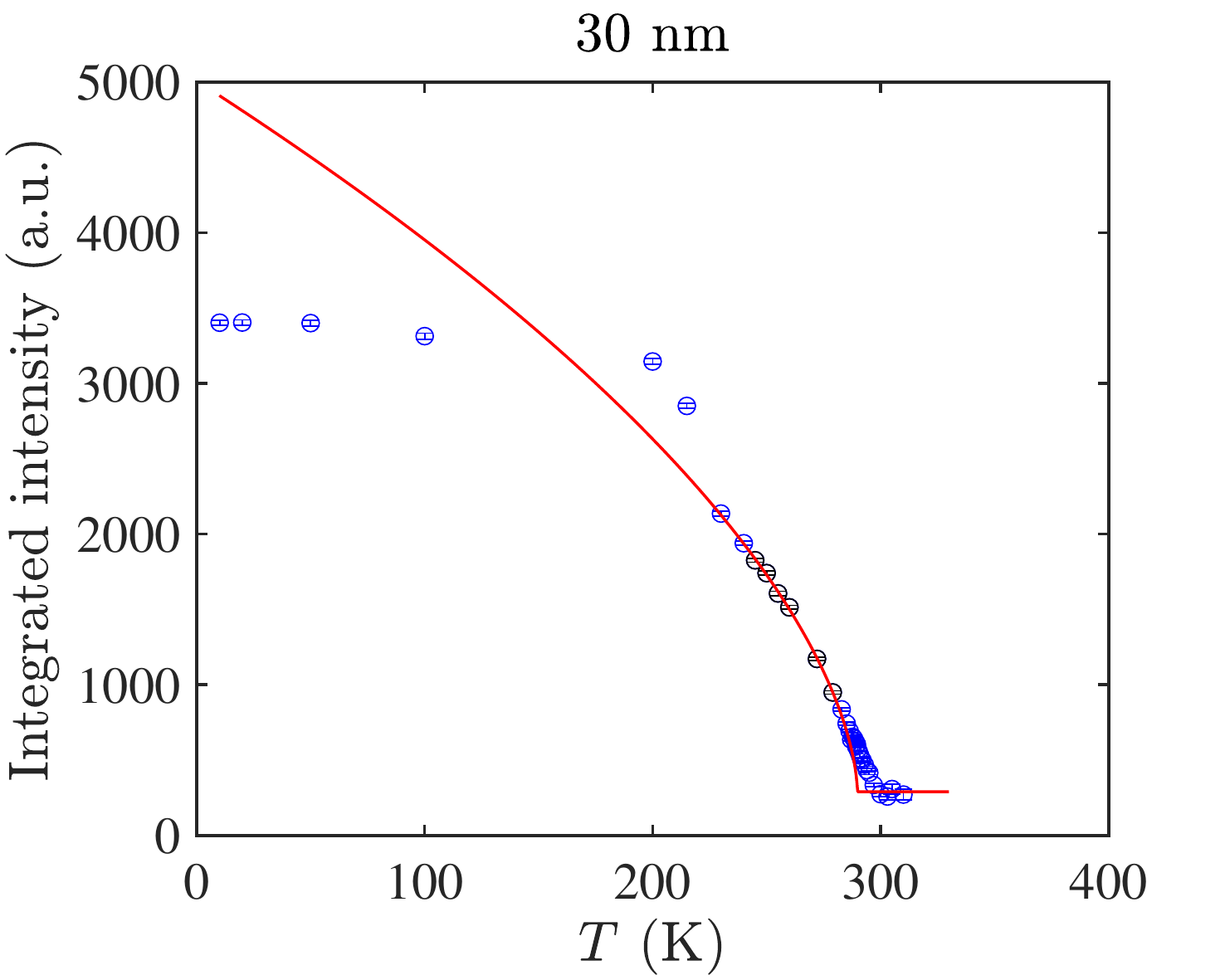}
\includegraphics[width=0.48\textwidth]{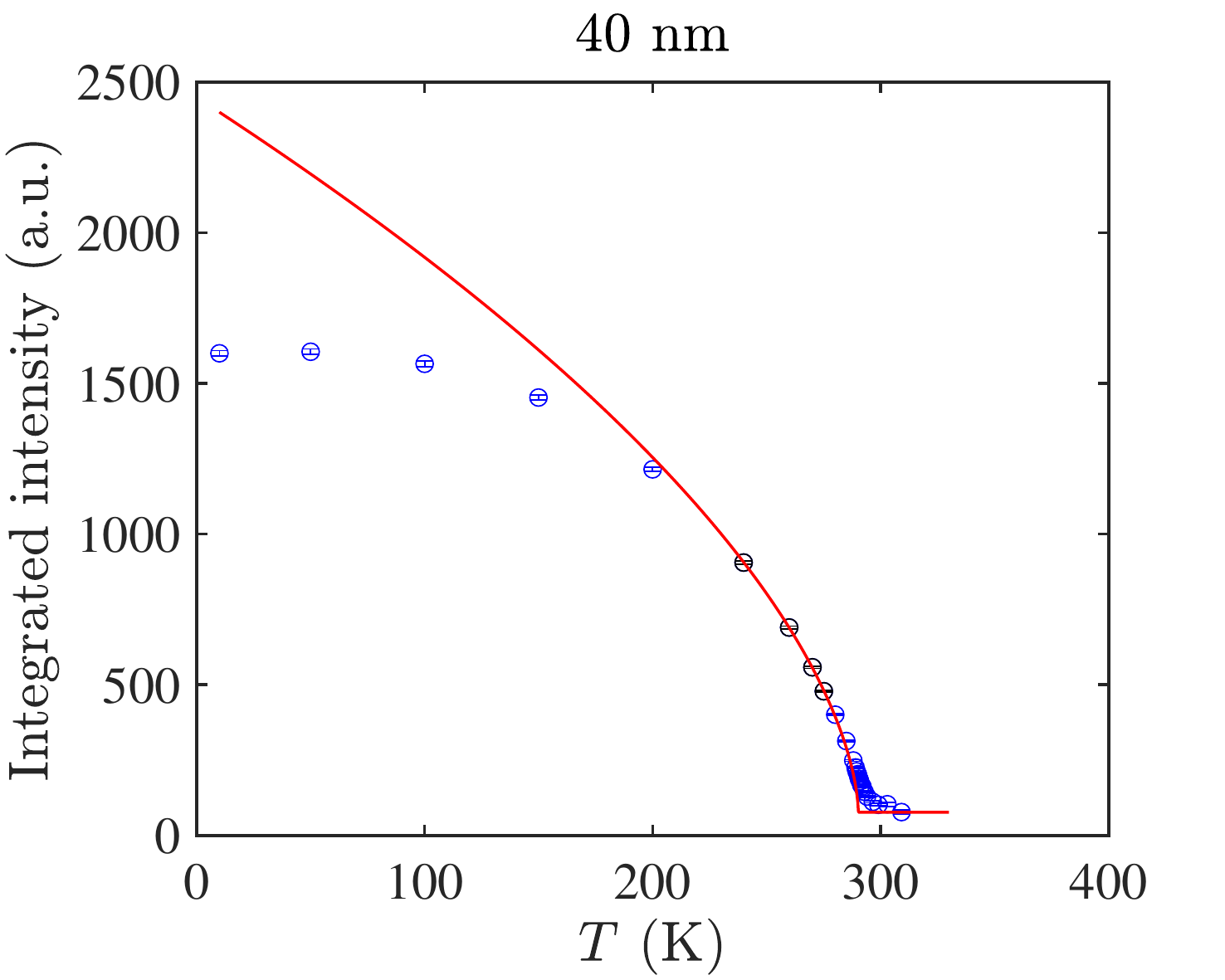}

\includegraphics[width=0.48\textwidth]{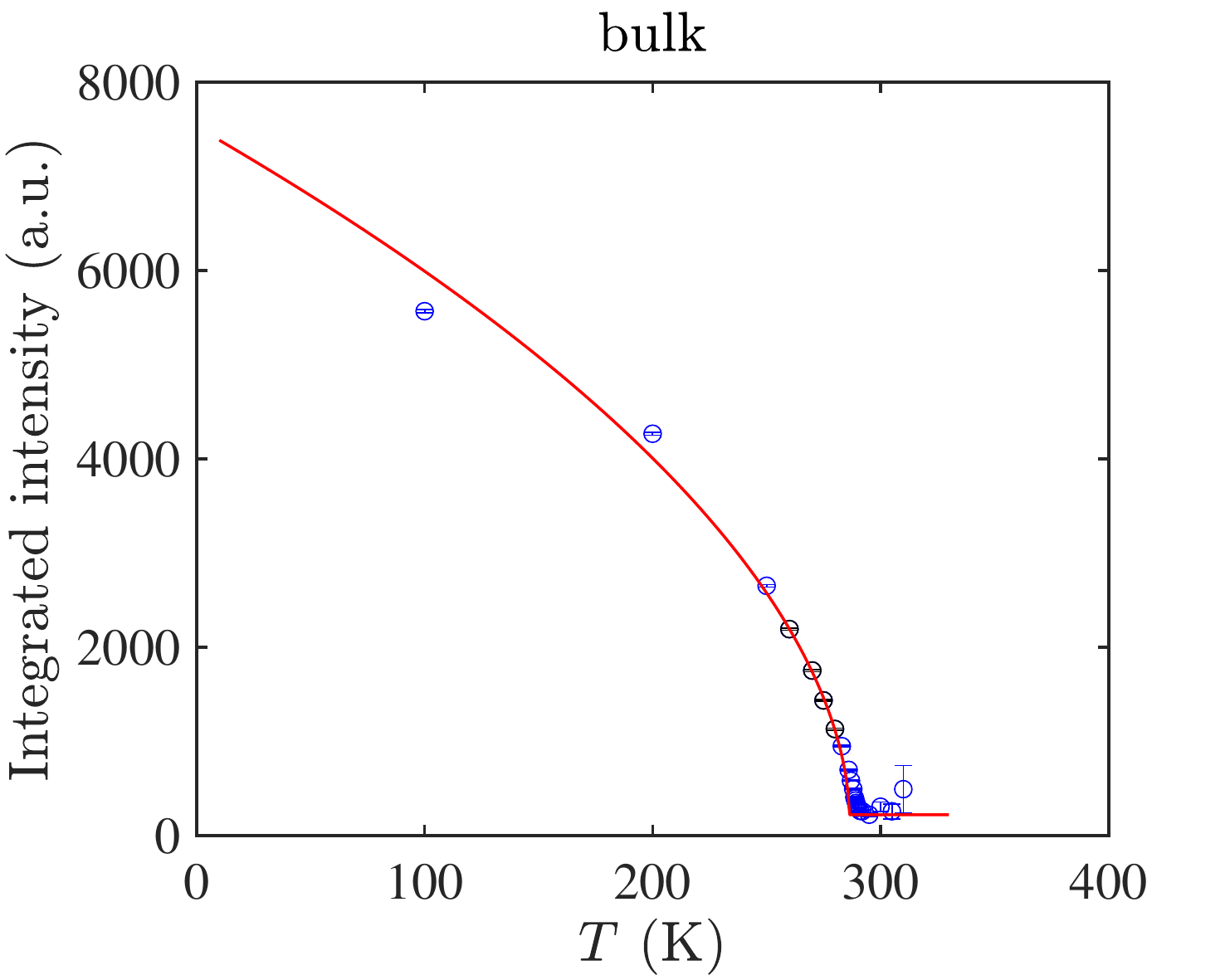}
\caption{Determination of the critical exponent $\beta$ for the antiferromagnetic order parameter, $M_{\rm AF} \propto (-t)^\beta$, by means of a power-law fit to the integrated neutron scattering intensity of the magnetic Bragg peaks for temperatures below $T_{\rm N}$. The red curves denote the fit for temperatures close to $T_{\rm N}$, incorporating only the black data points in the fitting routine. Note that $T_{\rm N}$ was determined by the least-squares method as explained in the main text. The values of $T_{\rm N}$ are 289.8(5) K, 290.2(4) K and 286.2(4) K for the 30 nm, 40 nm and bulk samples, respectively.
From the fitting, values of $2\beta$ were obtained, and these values are 0.598(13), 0.599(7) and 0.548(26), for the 30 nm, 40 nm and bulk samples, respectively. While these values are slightly lower than the Ising prediction of 0.625 which CoO should assume, they are close to values reported by Rechtin \textit{et. al.}, who argue that their smaller values were caused by a tetragonal lattice contraction below $T_{\rm N}$. \cite{Rechtin1970}}
\label{Fig:criticalexponent}
\end{figure}
%=================== FIGURE S4 ==================================

\clearpage
\section{Monte Carlo simulations}
The Monte Carlo simulations of the CoO nanoparticles use the nearest neighbour Ising model with spins polarized along the $z$-axis in the absence of external magnetic fields:

\begin{equation}
    H = -J\sum_{\langle i,j\rangle}s_is_j,
\end{equation}
where $J$ is the coupling strength, $s_i$ ($s_j$) is spin value in units of $1/2\hbar$ at the $i$'th ($j$'th) lattice position, and the brackets indicate that the sum runs over nearest neighbours. Using a simple cubic lattice, each lattice position has six nearest neighbours, except for those at the surface. Approximately spherical, monodisperse nanoparticles were simulated by only including lattice positions within a sphere with a diameter equal to the particle size. The lattice constant was set equal to the Co-Co distance in CoO; $a=4.2615/\sqrt(2)$ \AA{}. 
We used the Metropolis Hastings algorithm in our simulations.\cite{metropolis1953,hastings1970} In each step, a random lattice point is chosen, and the change in energy upon a spin flip is calculated. If the flip lowers the energy of the system, it is always performed, and if not, the flip is only carried out with a probability of $\exp[{-\Delta E/T_s}]$, where $\Delta E$ is the change in energy and $T_s$ is the simulated temperature of the system. For simplicity, we set $J= 1$. 
The Mersenne Twister algorithm was used to generate random numbers.\cite{matsumoto1998} 

The transition temperature of the system was found to be $T_{\rm N}\approx 4.5$. We therefore initiated each simulation at $T_s=6$, well above $T_N$. In this paramagnetic state, the spins were randomly set to either up or down. The temperature was then gradually lowered to $T_{\rm s} = 0$, and a simulation was carried out at each desired temperature, using the final spin configuration of the previous temperature as the starting point.

We found that the number of attempted spin flips, $N$, required for the simulations to converge, scales as $N \propto l^4$, where $l$ is the number of spins along the diameter of the particle. To ensure convergence, we set $N = 10 \ l^4$ in all simulations.

The simulation temperatures were chosen relative to the bulk transition temperature, $T_{\rm N}$ (which is around 4.5), in three different intervals: a linear distribution of sixteen temperatures between $T_{\rm s} = 6$ and $T_{\rm N}$, 185 temperatures distributed according to a power-law between $T_{\rm N}$ and $T_{\rm s} = 3$, and a linear distribution of five temperatures between $T_{\rm s} = 3$ and $T_{\rm s} = 0$. 

The neutron scattering intensity was calculated using: 
\begin{equation}
    I = \sum_{i,j} s_is_j\exp{i\bf{Q}\cdot\left(\bf{r}_j-\bf{r}_k\right)},
\end{equation}
where ${\bf Q}$ is the scattering vector and ${\bf r_i}$ and ${\bf r_j}$ are the positions of the $i$'th and $j$'th spin, respectively. The calculation can be simplified by noting that, since all sites are counted twice, the exponent can be reduced to a cosine summing only once over each lattice position.

In this work, we compare the scattering intensity from the simulations to powder-averaged experimental neutron data data. Performing the same powder average on spherically symmetric nanoparticles, only ${\bf{Q}}$ along one direction contributes to the measured signal. Thus, by setting ${\bf {Q}} = (Q,0,0)$, the scattering intensity simplifies to:

\begin{equation}
    I(Q) = \sum_i^{l^3}s_i\sum_k^lP_{\rm k}\cos{Q\left(x_i-x_k\right)},
\end{equation}
where $i$ runs over all spins and $k$ runs over all planes of spins orthogonal to ${\bf {Q}}$. In the $k$'th plane, the total spin is $P_{\rm k}$. This greatly reduces the evaluation of the scattering intensity from a sum over $l^6$ combinations to $l^4$. Finally, due to the symmetric setup of the simulations, only positive values of $Q$ were calculated. While this means that the peak in intensity occurs at $Q = 0$, an $x$-axis offset of around 1.28 \AA{} was introduced in Figure 3 of the main text to match the peak position of the simulations with that obtained for the neutron data.

All simulations were performed on the Quantum Wolf Cluster located at the Quantum Magnetism Department of \'Ecole Polytechnique F\'ed\'erale de Lausanne. The number of individually simulated system sizes was chosen to optimize the usage of the cluster, with the total number of simulations for each particle size tabulated in Table~\ref{tab:particleSizeSimNumber}.

\begin{table}[h!]
    \centering
    \begin{tabular}{c|c||c|c}
        \ Particle size $l$ \ & \ Number of simulations \ & \ Particle size $l$ \ & \ Number of simulations \ \\\hline
        19 & 311 & 71 & 310\\
        21 & 201 & 73 & 198\\
        23 & 203 & 75 & 201\\
        25 & 203 & 77 & 198\\
        27 & 139 & 79 & 203\\
        29 & 200 & 81 & 200\\
        31 & 201 & 83 & 200\\
        33 & 201 & 85 & 98\\
        35 & 204 & 87 & 107\\
        37 & 201 & 89 & 100\\
        39 & 201 & 91 & 132\\
        41 & 201 & 93 & 132\\
        43 & 200 & 95 & 99\\
        45 & 200 & 97 & 101\\
        47 & 198 & 99 & 98\\
        49 & 201 & 101 & 100\\
        51 & 200 & 103 & 100\\
        53 & 200 & 105 & 104\\
        55 & 198 & 107 & 50\\
        57 & 194 & 109 & 50\\
        59 & 201 & 111 & 50\\
        61 & 196 & 113 & 49\\
        63 & 306 & 115 & 50\\
        65 & 336 & 117 & 49\\
        67 & 237 & 119 & 50\\
        69 & 198 & 121 & 50\\\hline
    \end{tabular}
    \caption{Number of simulations performed for each particle size.}
    \label{tab:particleSizeSimNumber}
\end{table}

At zero temperature we find no evidence of any dead layer.

{}